\documentclass[aps,physrev,onecolumn,groupedaddress]{revtex4-2}
\PassOptionsToPackage{breaklinks=true}{hyperref}
\AtBeginDocument{%
  \captionsetup{justification=justified,singlelinecheck=false}%
  \captionsetup[sub]{justification=justified,singlelinecheck=false}%
}
\usepackage{graphicx}
\usepackage{amsmath}
\usepackage[compatibility=false]{caption}
\usepackage{subcaption}
\usepackage{orcidlink}

\usepackage{hyperref} 

\usepackage[usenames,dvipsnames]{xcolor}
\usepackage{color}

\newcommand{\pd}[4]{\frac{\partial #1 ^{#2}}{\partial #3^{#4}}}
\newcommand{\p}[1]{\left( #1 \right)}

\newcommand{\bra}[1]{\left\langle #1 \right\vert}
\newcommand{\ket}[1]{\left\vert #1 \right\rangle}
\newcommand{\braket}[2]{\left\langle #1 \middle| #2 \right\rangle}
\newcommand{\braketmid}[3]{\left\langle #1 \middle| #2 \middle| #3 \right \rangle}

\newcommand\myshade{85}
\colorlet{mylinkcolor}{violet}
\colorlet{mycitecolor}{YellowOrange}
\colorlet{myurlcolor}{Aquamarine}
\hypersetup{
  breaklinks = true,
  linkcolor  = mylinkcolor!\myshade!black,
  citecolor  = mycitecolor!\myshade!black,
  urlcolor   = myurlcolor!\myshade!black,
  colorlinks = true,
}

\begin{document}

\title{Designing quantum technologies with a quantum computer}

\author{Juan Naranjo$^1$ \orcidlink{0009-0007-4456-7083}}
\affiliation{$^1$LG Electronics Toronto AI Lab}

\author{Thi Ha Kyaw$^1$ \orcidlink{0000-0002-3557-2709}}
\email[]{thiha.kyaw@lge.com}
\affiliation{$^1$LG Electronics Toronto AI Lab}

\author{Gaurav Saxena$^1$ \orcidlink{0000-0001-6551-1782}}
\affiliation{$^1$LG Electronics Toronto AI Lab}

\author{Kevin Ferreira$^1$}
\affiliation{$^1$LG Electronics Toronto AI Lab}

\author{Jack S. Baker$^1$ \orcidlink{0000-0001-6635-1397}}
\email[]{jack.baker@lge.com}
\affiliation{$^1$LG Electronics Toronto AI Lab}

\date{\today}

\begin{abstract}
Interacting spin systems in solids underpin a wide range of quantum technologies, from quantum sensors and single-photon sources to spin-defect-based quantum registers and processors. We develop a quantum-computer-aided framework for simulating such devices using a general many-body electron-spin-resonance Hamiltonian that incorporates zero-field splitting, the Zeeman effect, hyperfine interactions, dipole-dipole spin-spin interactions, and electron-phonon decoherence. Within this framework, we combine Gray-encoded qudit-to-qubit mappings, qubit-wise commuting aggregation, and a multi-reference selected quantum Krylov fast-forwarding hybrid algorithm, aiming to access extended-time dynamics within the constraints of NISQ and early fault-tolerant hardware. Numerical simulations demonstrate the computation of operationally useful quantities including autocorrelation functions up to $\sim100$ ns, together with microwave absorption spectra and the $\ell_1$-norm of coherence, achieving 18-30$\%$ reductions in gate counts and circuit depth for Trotterized time-evolution circuits compared to unoptimized implementations. Using the nitrogen vacancy center in diamond as a testbed, we benchmark the framework against classical simulations and identify the reference-state selection in sQKFF as the primary factor governing accuracy at fixed hardware cost. This methodology provides a flexible blueprint for using quantum computers to design, compare, and optimize solid-state spin-qubit technologies under experimentally realistic conditions.
\end{abstract}

\maketitle

\section{Introduction \label{sec:intro}}
Solid-state materials embedded with spin-active defects are increasingly attracting attention for applications as quantum sensors, single-photon sources, and quantum processors. The search for new spin defects and suitable host materials is an active area of research, where candidate systems must allow for precise manipulation and readout and permit long coherence times \cite{Maciaszek2025, Vogl2025, Rubinas2025, Carter2025}. Among these systems, nitrogen-vacancy (NV) centers have emerged as a leading platform due to their long coherence, dephasing, and relaxation times, which can be further extended through techniques such as Ramsey and Hahn echo sequences \cite{Barry2020, NV_2}. Nevertheless, the complex spin–environment interactions and internal spin couplings pose significant challenges for classical modeling and analysis \cite{Park2022, Suter2017}. Digital quantum computer-aided design \cite{Kyaw2021Oct, Kottmann2021Aug} offers a promising alternative, giving access to the dynamics of these systems under realistic conditions and to the derived quantities that follow from them, such as the coherence times of arbitrary states and microwave absorption spectra.

\begin{figure*}
    \centering
    \includegraphics[width=\linewidth]{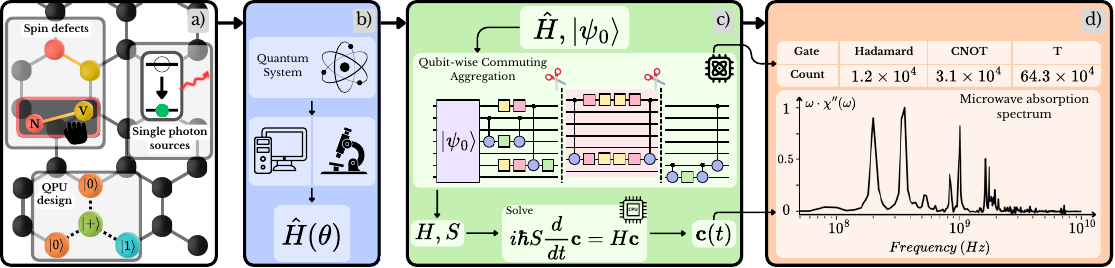}
    \caption{The quantum computer-aided design framework proposed in this work. The process begins by defining a quantum system (a) by specifying spin-defect species, the host material, the presence of nuclear spin species, applied magnetic fields, and the geometry of the spin ensemble (configurations 1, 2, and 3 in Table \ref{tab:TableParameters} are examples of such a specification). Next, the system Hamiltonian is constructed, with its parameters obtained either computationally or experimentally, as depicted in (b). With the Hamiltonian defined, the sQKFF algorithm can be executed by specifying the system’s initial state. (c) is divided into two parts: the quantum component, which computes the elements of the $H$ and $S$ matrices via Hadamard tests and performs the optimization step through QWC aggregation; and the classical component, which solves the Schrödinger equation within the Krylov subspace. Finally, as shown in (d), the outputs of (c) enable the estimation of quantum resource requirements and the computation of key system properties such as the system’s autocorrelation function, microwave absorption spectrum, and time-dependent $\ell_1$-norm of coherence of other dynamics-derived properties.}
    \label{fig:HeroFigure}
\end{figure*}

More broadly, the simulation of quantum systems \cite{Georgescu2014} stands as one of the most promising applications of quantum computing, alongside drug discovery \cite{DrugDisc_1, DrugDisc_2}, communications \cite{Comms_1, Comms_2}, optimization \cite{Energy_1, Energy_2}, machine learning \cite{ML_2, ML_3, ML_4}, and cryptography \cite{Foreman2023, Qasim2026}. Nevertheless, current quantum technologies still operate within the Noisy Intermediate-Scale Quantum (NISQ) era, transitioning toward early fault-tolerant devices, where limited resources and intrinsic noise constrain hardware \cite{Preskill2018, Katabarwa2023}. Therefore, a central research direction is the development of efficient algorithms that can approximate the time evolution of quantum systems while minimizing the required computational resources. Key optimization metrics include the number of single- and multi-qubit gates, qubit counts, and circuit depth, thereby mitigating restrictions imposed by various sources of noise \cite{Childs2018, Brown2010}.

Spin Hamiltonians serve as coarse-grained models in chemistry and materials science, capturing collective magnetic interactions in a simplified form. When calibrated with accurate parameters, these effective models reproduce complex phenomena, yet they remain many-body in nature and are generally intractable for classical computation. A prominent example is the nuclear magnetic resonance (NMR) spin Hamiltonian, which has become a focus of recent quantum-simulation efforts on gate-based processors. Google Quantum AI and collaborators recently provided evidence that simulating higher-order out-of-time-ordered correlators (OTOCs) of an NMR Hamiltonian lies beyond the reach of state-of-the-art classical methods within practical runtimes \cite{Google2025}; in subsequent work, OTOC data measured in many-body nuclear spin-echo experiments were used to train a parameterized model and infer molecular geometric information \cite{Zhang2025, OBrien2022}. These NMR studies bear directly on the present work, because the NMR spin Hamiltonian is a sub-model of the more general electron-spin-resonance (ESR) Hamiltonian we consider, which further couples electronic and nuclear spin degrees of freedom and adds a spin-phonon interaction characteristic of solid-state defects. The simulation problem we target is therefore at least as demanding as the NMR case already shown to be hard.

The ESR spin-defect models considered here describe a collection of interacting spin defects embedded in a solid with vibrational degrees of freedom. This model class underpins platforms such as the NV center, yet it has received comparatively little attention in the quantum-simulation literature relative to its technological importance. The few existing quantum-algorithmic studies of spin defects in solids \cite{Huang2022, Baker2024, Casares2025} typically target the underlying electronic-structure problem, either in full or in a reduced form obtained through quantum embedding \cite{Sheng2022, Chen2025}, at considerable quantum-resource cost. We take a complementary route: we simulate the ESR Hamiltonian with an effective spin-boson coupling directly, rather than reconstructing it from a first-principles electron-phonon treatment. This retains the physics that governs experimentally observable spin dynamics while keeping the resource requirements compatible with near-term hardware, at the cost of adopting an effective, pre-parameterized model, as discussed in Section~\ref{sec:methods}.

In this work we present a framework for simulating and characterizing spin defects in solid-state materials, with the aim of showing how realistic ESR-type Hamiltonians can be integrated into hybrid quantum-computing workflows. As illustrated in figure \ref{fig:HeroFigure}, the methodology combines several resource-aware techniques compatible with current quantum hardware and is benchmarked on three representative NV$^{-}$ center configurations. The procedure begins by defining a spin-defect system in a given host material, whose Hamiltonian parameters are taken from experiment or computed classically. The Hamiltonian is then encoded into a spin-$1/2$ representation for time-evolution simulation and the computation of relevant properties.

Operationally relevant quantities, such as coherence and dephasing times, often emerge over comparatively long timescales, from nanoseconds to microseconds \cite{Barry2020}. Resolving them by direct Trotter–Suzuki product-formula evolution is computationally infeasible, as the circuit depth grows with the simulated time \cite{PF_Trotter_Error_Childs2019}. Hybrid algorithms address this limitation by fast-forwarding the time evolution at reduced cost \cite{Gu2021, Crstoiu2020, Commeau2020, Shang2025, Ko2025}. We adopt the multi-reference selected Quantum Krylov Fast-Forwarding (sQKFF) algorithm \cite{sQKFF_Algorithm}, a physically motivated heuristic in which the accessible evolution time is found to depend on the choice and number of reference states used to span the relevant dynamical subspace. We further incorporate an optimization stage based on qubit-wise commuting (QWC) aggregation \cite{Verteletskyi2019, JoshIzaac2021}, which reduces the gate count and depth of the Trotterized circuits.

The remainder of this work is organized as follows. Section \ref{sec:methods} provides a detailed explanation of each step of the proposed framework, including key methodological considerations and descriptions of the NV$^{-}$ configurations used in numerical simulation. Section \ref{sec:results} presents the results, focusing on the comparison between properties approximated using both the sQKFF algorithm and QuTiP, including an analysis of the influence of varying the Trotterization error and the number of reference states, $R$, in the sQKFF algorithm. Resource estimations are presented in this section. Section \ref{sec:consid_and_outlook} discusses additional factors that may affect the results and outlines potential directions for future research. Finally, Section \ref{sec:summary_and_conclusions} summarizes the main findings and conclusions of this work.

\begin{table*}[!t]
    \centering
    {\small
    \begin{tabular}{c|lll}
    \hline
    \textbf{Configuration} & \multicolumn{1}{c|}{\textbf{1}} & \multicolumn{1}{c|}{\textbf{2}} & \multicolumn{1}{c}{\textbf{3}} \\ \hline
    \textbf{\begin{tabular}[c]{@{}c@{}}Specific\\parameters\end{tabular}} & \multicolumn{1}{l|}{\begin{tabular}[c]{@{}l@{}}- 1 NV$^{-}$ centre\\ - $\ket{\varphi} = \frac{1}{\sqrt{2}}\p{\ket{0} + \ket{2}}$ \end{tabular}} & \multicolumn{1}{l|}{\begin{tabular}[c]{@{}l@{}}- 3 NV$^{-}$ centres\\ - $\ket{\varphi} = \frac{1}{\sqrt{2}}\p{\ket{000} + \ket{222}}$ \end{tabular}}   & \begin{tabular}[c]{@{}l@{}}- 1 NV$^{-}$ centre\\ - 2 $N^{14}$ impurities\\ - $\ket{\varphi} = \frac{1}{\sqrt{2}}\p{\ket{0} + \ket{2}}$ \end{tabular}   \\ \hline
    \textbf{\begin{tabular}[c]{@{}c@{}}Common\\parameters\end{tabular}}     & \multicolumn{3}{l}{\begin{tabular}[c]{@{}l@{}}- Magnetic field of 2 mT in the $z$ direction\\ - Boson bath of dimension 8, with a single frequency mode of 5.8 GHz\\ - Spin-boson coupling intensity of 1.78 GHz and coupling operator $S_{x}$ \\ - Initial state of the bath: $Ry\p{\frac{\pi}{2}}\otimes Ry\p{\frac{\pi}{4}} \otimes Ry\p{\frac{\pi}{8}}\ket{000}$\\ - Initial state of each NV$^{-}$ nuclear spin and/or $N^{14}$ impurity: $\ket{0}$ \end{tabular}} \\ \hline
    \end{tabular}
    }
    \caption{Parameters of the three NV$^{-}$ centre configurations and their interactions. $\ket{\varphi}$ denotes the initial state of the electronic spin of the NV$^{-}$ centres, with $\ket{2}\equiv\ket{m_{s}= +1}$. In configurations 2 and 3, the three components, NV$^{-}$s and/or $N^{14}$ impurities, are arranged at the vertices of an equilateral triangle with a side length of 1 nm.}
    \label{tab:TableParameters}
\end{table*}

\section{Methods \label{sec:methods}} 
\subsection{Hamiltonian Models for Spin-Defects Simulations.}
The ESR Hamiltonian employed in this work should be understood as a low-energy description of the spin-defect system. Its construction relies on the projection of the underlying electronic structure onto the relevant spin degrees of freedom, retaining the interactions that govern experimentally observable spin dynamics, such as zero-field splitting, Zeeman, hyperfine, dipolar, and spin-phonon coupling terms. As a consequence, higher-order effects, multi-mode environmental interactions, charge-noise processes, and other microscopic degrees of freedom are not explicitly included. Despite these approximations, effective spin Hamiltonians of this form constitute the canonical framework for describing ESR and EPR experiments, and have been extensively validated in the study of NV centers and other solid-state spin defects \cite{Suter2017, Kong2018, Karjalainen2025}.

Within a proof-of-concept setting aimed at supporting the prototyping of spin defects in solid-state materials, the framework incorporates the interactions mentioned in the previous paragraph, which are operational and environmental ones. However, recent lines of investigation are moving towards designing Hamiltonians of materials with desired properties \cite{Kokail2026}, which, if expressed in terms of Pauli strings, can also be simulated with the present framework. A detailed discussion of the Hamiltonian terms used in this work can be found in \cite{Barry2020, Park2022, Wang2024}.

Although the present study focuses on interacting NV$^{-}$ spin defects, the framework itself is not restricted to a particular defect platform. More generally, it accepts spin Hamiltonians that can be expressed as sums of Pauli operators after an appropriate encoding. For spin-$1/2$ defect systems, such as donor spins in silicon or several color-center platforms, the encoding is particularly straightforward since each spin can be represented directly by a single qubit. Higher-spin defects can likewise be accommodated through larger Gray-encoded registers, with the required number of qubits determined by the local spin dimension \cite{Sawaya2020}. Hence, the Hamiltonian decomposition, commuting-term aggregation, and Krylov-based time-forwarding stages of the workflow remain applicable across a broad class of spin-defect systems, with only the Hamiltonian parameters and local encoding requiring modification. Consequently, the present implementation is taken primarily as a representative case study of a framework thought to be widely applicable.

To describe the dynamics of a system with $\mathcal{N}$ NV$^-$ centers, the zero-field splitting and Zeeman terms,
\begin{equation}
    \label{ec:NV_H_ZeroField_Zeeman}
    \hat{H}_{S} = \hbar D\sum_{i=1}^{\mathcal{N}}\hat{S}_{z,i}^{2} + g_{NV}\mu_{B} \vec{B}\cdot\sum_{i=1}^{\mathcal{N}}\vec{S_i},
\end{equation}

\noindent are included. These describe the electronic spins' interaction with an external magnetic field $\vec{B}$. Here, $D = 2.87$ GHz denotes the zero-field splitting parameter, $g_{NV} = 2.003$ the NV g-factor, $\mu_B$ the Bohr magneton, and $\vec{S_i} = [S_{x,i}, S_{y,i}, S_{z,i}]$ the electronic spin-1 operator of the $i$-th NV$^{-}$ center \cite{Barry2020, Felton2008}.

The hyperfine interaction between the electronic spin and the $^{14}N$ nuclear spin of the NV$^{-}$ centers is described by
\begin{equation}
    \label{ec:NV_H_14N_nuclei}
    \begin{split}
        \hat{H}_{N} = {} & g_{N}\mu_{N}\vec{B}\cdot\sum_{i=1}^{\mathcal{N}}\vec{I_i}
        + \hbar\sum_{i=1}^{\mathcal{N}}\Big[ A_{\|}\hat{S}_{z,i}\hat{I}_{z,i} \\
        & + A_{\perp}\p{\hat{S}_{x,i}\hat{I}_{x,i}+\hat{S}_{y,i}\hat{I}_{y,i}}
        + Q\p{\hat{I}_{z,i}^{2} - \frac{2}{3}\hat{1}}\Big],
    \end{split}
\end{equation}

\noindent with $g_N = 0.403573$ the nuclear g-factor, $\mu_N$ the nuclear magneton, $A_{\|}=-2.16$ MHz and $A_{\perp}=-2.7$ MHz the axial and transverse hyperfine coupling constants, $Q = 5$ MHz the nuclear electric quadrupole parameter, and $\vec{I_i} = [I_{x,i}, I_{y,i}, I_{z,i}]$ the spin-1 operator of the $i$-th NV$^{-}$ center's $^{14}N$ nucleus \cite{Barry2020, Felton2008, Steiner2010}. Note that the Zeeman and quadrupole terms in equation \ref{ec:NV_H_14N_nuclei} correspond to the dynamics of a $^{14}N$ impurity within the lattice.

One of the primary sources of decoherence in spin defects arises from coupling to lattice vibrations. The framework incorporates this interaction through the spin–boson Hamiltonian
\begin{equation}
    \label{ec:boson_hamiltonian}
    \hat{H}_{B} = \hbar\omega\hat{b}^{\dagger}\hat{b} + \hbar\lambda \sum_{i=1}^{\mathcal{N}}\hat{S}_{x,i}\otimes\p{\hat{b}^{\dagger} + \hat{b}},
\end{equation}

\noindent where $\omega$ is the frequency of the phonon mode, $\lambda$ is the coupling strength, and $b$, $b^{\dagger}$ are the bosonic annihilation and creation operators, respectively. Equation \ref{ec:boson_hamiltonian} models a simplified single vibrational mode collectively coupled to the NV$^{-}$ centers within a truncated bosonic description. This approximation is intended as a proof-of-principle representation of spin-phonon coupling effects and does not capture more realistic environmental processes such as multi-mode phonon baths, non-Markovian dynamics, or charge-noise-induced decoherence \cite{Ganhari2026, Candido2024}. However, extensions to multiple bosonic modes and more selective coupling structures are, in principle, possible within the same Hamiltonian decomposition framework, although such cases were not explored in the present work. Nonetheless, the coupling strength considered in this work is comparable in magnitude to characteristic spin energy scales such as the NV$^{-}$ zero-field splitting, see table \ref{tab:TableParameters}, placing the model in a regime where spin-phonon interactions can significantly influence the dynamics.

Additional decoherence sources include proximal electronic spins from other NV$^{-}$ centers and nuclear spins from $^{14}$N impurities. This interaction is captured via the dipole–dipole Hamiltonian, described by
\begin{equation}
    \label{ec:dipolar_coupling_hamiltonian}
    \hat{H}_{\text{dd}} = \frac{\mu_{0}\alpha_{1}\alpha_{2}}{4\pi r^{3}}\left[\vec{S}_{1}\cdot\vec{S}_{2} - 3\p{\vec{S}_{1}\cdot\hat{r}}\p{\vec{S}_{2}\cdot\hat{r}}\right],
\end{equation}

\noindent with $\mu_0$ the vacuum permeability; then, for each spin 1 and 2: $\alpha$ the product of their respective g-factor and magneton, and $\vec{S}$ is their spin operator. These spins are separated by a distance $r$, with $\hat{r}$ the unit direction vector.

\begin{figure*}
    \centering
    \includegraphics[width=0.8\linewidth]{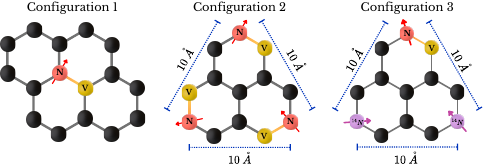}
\caption{Spatial layout of the three benchmark NV$^{-}$ configurations studied in this work, which differ in the number of spins and the strength of their interactions. NV$^{-}$ centers are shown as yellow and red elements and $^{14}$N impurities as purple elements. Together with the parameters in table \ref{tab:TableParameters}, these geometries fix the dipole--dipole and hyperfine couplings entering the Hamiltonian; Configurations 2 and 3 place their spins on an equilateral triangle of side $1$ nm.}
    \label{fig:Configurations_layout}
\end{figure*}

As a testbed to analyze the role of individual interactions and to benchmark our method, we simulate three representative NV$^{-}$ configurations with their parameters summarized in table \ref{tab:TableParameters} and their layouts sketched in figure \ref{fig:Configurations_layout}. For efficient implementation, the relevant $d$-level operators (i.e. the spin-$S$ and truncated bosonic operators) are mapped to a spin-$1/2$ representation via Gray encoding, requiring $\left\lceil log_{2}(d) \right\rceil$ qubits \cite{dlevelSim}.

\subsection{Qubit-wise Commuting Partitioning and Circuit Optimization}
The Hamiltonian $\hat{H} = \hat{H}_S + \hat{H}_N + \hat{H}_B + \hat{H}_{\text{dd}}$, decomposed into a linear combination of Pauli strings, is partitioned into qubit-wise commuting (QWC) groups, whose terms commute within each qubit subspace and can therefore be simultaneously diagonalized via single-qubit Clifford rotations \cite{Verteletskyi2019, JoshIzaac2021}. This structure simplifies Trotterized time evolution by enabling canonical exponentiation circuits for diagonal Pauli operators \cite{dlevelSim}. Moreover, grouping permits optimal ordering of exponentials, allowing cancellation of adjacent adjoint operations and reducing both gate count and circuit depth.

\subsection{The sQKFF algorithm}
\subsubsection{Implementation via product-formula simulation.}
Given an efficient approximation of the time-evolution operator, the sQKFF algorithm represents the evolved state within a Krylov subspace as
\begin{equation}
    \label{ec:krylov_approx}
    \ket{\psi(t)} = e^{-it\hat{H}/\hbar}\ket{\psi_{0}} \approx \sum_{m=0}^{M-1}\sum_{r=1}^{R}c_{mr}(t)\ket{\phi_{mr}},
\end{equation}
with $\ket{\psi_{0}}$ the initial state, $c_{mr}(t)$ time-dependent complex coefficients, and $\ket{\phi_{mr}}= e^{-im\tau\hat{H}}\ket{r}$. This construction spans an $M$-order Krylov subspace generated from $R$ reference states $\ket{r} \in \{\ket{\psi_0},\ket{\mathcal{B}_{1}},\cdots,\ket{\mathcal{B}_{R-1}}\}$. The auxiliary states $\ket{\mathcal{B}_{r}}$ are bitstrings sampled from measurements of $e^{-i(M-1)\tau \hat{H}}\ket{\psi_0}$ and retained according to their outcome probabilities.

This reference-state selection follows the physically motivated heuristic of Ref.~\cite{sQKFF_Algorithm}. From a Krylov-subspace standpoint, convergence is governed by how well the constructed subspace captures the spectral components that dominate the dynamics of the initial state \cite{sQK_Algorithm}; high-probability bitstrings are therefore expected to bias the subspace toward the dynamically relevant sector of Hilbert space, yielding a more informative-reduced representation of the evolved state. Unlike conventional Krylov constructions based on repeated applications of $\hat{H}$, however, this strategy is heuristic and carries no \textit{a priori} optimality or convergence guarantee; the long-time accuracy of the approximation consequently depends strongly on the quality and number of the selected reference states.

Substituting equation~\ref{ec:krylov_approx} into the Schr\"odinger equation and projecting onto $\bra{\phi_{m'r'}}$ yields the generalized linear system
\begin{equation}
    \label{ec:Krylov_Schrodinger_eq}
    i\hbar S\pd{}{}{t}{}\textbf{c} = H\textbf{c},
\end{equation}
where $S$ and $H$ are $MR\times MR$ matrices with elements $\braket{\phi_{m'r'}}{\phi_{mr}}$ and $\braketmid{\phi_{m'r'}}{\hat{H}}{\phi_{mr}}$, respectively, and $\mathbf{c}$ is an $MR\times 1$ coefficient vector. Solving equation~\ref{ec:Krylov_Schrodinger_eq} constitutes the classical component of the algorithm, whereas the quantum component consists of estimating the elements of $S$ and $H$.

\begin{figure*}
    \centering
    \includegraphics[width=0.8\linewidth]{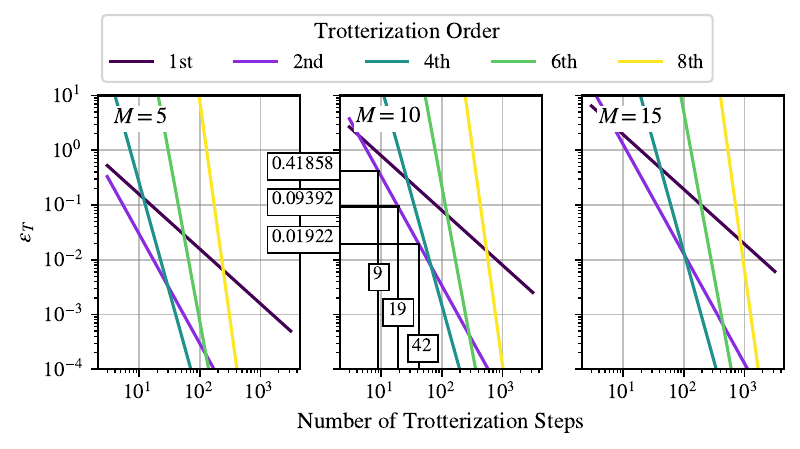}
\caption{The Trotterization error sets the circuit depth used in our simulations. Shown is the upper bound on the product-formula error $\varepsilon_{T}$ of equation \ref{ec:Trotter_error}, expressed through the Hamiltonian 1-norm $\|\hat{H}\|_{1}$, as a function of the number of Trotterization steps $s$, the product-formula order $k$, and the Krylov order $M$. Increasing $s$ or $k$ lowers $\varepsilon_{T}$; the three thresholds $\varepsilon_{T}=0.41858$, $0.09392$, and $0.01922$ used throughout this work correspond to $s=9$, $19$, and $42$ steps at $k=2$ and $M=10$.}
    \label{fig:TrotterError}
\end{figure*}

Evaluating these elements requires the propagators $e^{-im\tau\hat{H}}$, which we approximate using a $k$-th order product formula. Cort\'es and Gray \cite{sQK_Algorithm} bound the Krylov step from the spectral range $\Delta E$ of the Hamiltonian, $\tau\leq\pi\hbar/\Delta E$; using $\|\hat{H}\|_{1}\geq \Delta E/2$, with $\|\hat{H}\|_{1}$ the Hamiltonian 1-norm, we adopt the tighter choice $\tau\leq\pi\hbar/(2\|\hat{H}\|_{1})$. The associated product-formula (Trotterization) error over an evolution time $t$ then obeys the $\|\hat{H}\|_1$-dependent bound \cite{PF_Trotter_Error_Childs2019}
\begin{equation}
    \label{ec:Trotter_error}
    \varepsilon_{T} \leq \frac{1}{s^{k}}\cdot\frac{\p{t\|\hat{H}\|_1\hbar^{-1}}^{k+1}}{(k+1)!}\left[ \p{2\cdot5^{\frac{k}{2}-1}}^{k+1}+ 1 \right],
\end{equation}
where $s$ is the number of Trotterization steps and $k$ the product-formula order. With $t$ an integer multiple of $\tau$ up to $(M-1)\tau$, the choice $\tau\propto\|\hat{H}\|_{1}^{-1}$ renders $\varepsilon_{T}$ dependent only on $M$, $s$, and $k$. Accordingly, we fix $\tau = \frac{\pi\hbar}{10\|\hat{H}\|_{1}}$, $M=10$, and $k=2$, and determine the number of Trotterization steps required for $\varepsilon_{T}$ at $t=(M-1)\tau$ to fall below $0.5$, $0.1$, and $0.02$. These thresholds correspond to $9$, $19$, and $42$ steps, with exact values $\varepsilon_T = 0.41858$, $0.09392$, and $0.01922$, respectively, as shown in figure~\ref{fig:TrotterError}.

In the idealized continuous-time limit, the exact propagator commutes with $\hat{H}$ and satisfies $e^{it'\hat{H}}e^{-it\hat{H}}=e^{-i\p{t-t'}\hat{H}}$, so each element of $S$ and $H$ depends only on the index difference $m-m'$. Both matrices are therefore Toeplitz for every pair of reference states, and a single propagator per diagonal suffices to populate them. Under product-formula evolution this structure is only partially preserved. The overlap matrix $S$ remains \emph{exactly} Toeplitz, because powers of the approximate propagator compose without error, $\hat{V}_{m'}^{\dagger}\hat{V}_{m}=\hat{V}_{m-m'}$ with $\hat{V}_{j}\equiv(\hat{V}_{1})^{j}$. The Hamiltonian matrix $H$, by contrast, does not: $\hat{V}_{m'}^{\dagger}\hat{H}\hat{V}_{m}$ differs from the single-propagator form $\hat{H}\hat{V}_{m-m'}$ because the approximate propagator does not leave $\hat{H}$ invariant under conjugation.

We quantify this deviation directly in terms of $\varepsilon_{T}$. Let $\hat{U}_{j}=e^{-ij\tau\hat{H}}$ denote the exact propagator and $\hat{V}_{1}=\hat{U}_{1}+\mathcal{O}(\tau^{k+1})$ its $k$-th order product-formula approximation, with single-step error $\varepsilon_{1}\equiv\|\hat{V}_{1}-\hat{U}_{1}\|$ and the standard accumulation $\|\hat{V}_{j}-\hat{U}_{j}\|\leq j\varepsilon_{1}$. For indices $m\geq m'$, the non-Toeplitz and Toeplitz constructions of the Hamiltonian element differ by the operator
\begin{equation}
    \Delta_{m',m} \equiv \hat{V}_{m'}^{\dagger}\hat{H}\hat{V}_{m} - \hat{H}\hat{V}_{m-m'},
\end{equation}
and both forms approximate the common exact operator $\hat{U}_{m'}^{\dagger}\hat{H}\hat{U}_{m}=\hat{H}\hat{U}_{m-m'}$, the two exact expressions coinciding because $\hat{U}_{j}$ commutes with $\hat{H}$. Triangulating $\Delta_{m',m}$ through this common value and using $\|\hat{H}\|\leq\|\hat{H}\|_{1}$,
\begin{equation}
    \label{ec:toeplitz_legs}
    \begin{split}
        \|\Delta_{m',m}\| & \leq \|\hat{H}\|_{1}\Big(\|\hat{V}_{m'}-\hat{U}_{m'}\| + \|\hat{V}_{m}-\hat{U}_{m}\| \\
        & \qquad + \|\hat{V}_{m-m'}-\hat{U}_{m-m'}\|\Big) \leq 2m\,\|\hat{H}\|_{1}\,\varepsilon_{1},
    \end{split}
\end{equation}
where the first two terms account for the two independently approximated propagators of the non-Toeplitz element and the third for the single propagator of its Toeplitz surrogate. Since $|H^{\mathrm{NT}}_{m'r',mr} - H^{\mathrm{T}}_{m'r',mr}|\leq\|\Delta_{m',m}\|$ with $m\leq M-1$, and since the single-step bound $\varepsilon_{1}$ is smaller than the full-window bound $\varepsilon_{T}$ of equation~\ref{ec:Trotter_error} by a factor $(M-1)$ --- a direct consequence of the $t^{k+1}/s^{k}$ scaling for steps distributed uniformly over the window --- the largest element-wise deviation is simply
\begin{equation}
    \label{ec:toeplitz_bound}
    \max_{m,m',r,r'}\big|H^{\mathrm{NT}}_{m'r',mr} - H^{\mathrm{T}}_{m'r',mr}\big| \;\leq\; 2\,\|\hat{H}\|_{1}\,\varepsilon_{T}.
\end{equation}
The Toeplitz deviation thus inherits the magnitude of the Trotterization error itself, is largest for elements spanning the full window, and vanishes identically whenever either reference state is unevolved, where the two constructions coincide.

Two features of equation~\ref{ec:toeplitz_bound} clarify the role of the approximation. First, the product-formula error is relocated rather than removed. With $\hat{V}_{m}$ in place of the exact $\hat{U}_{m}$, the prepared states $\hat{V}_{m}\ket{r}$ differ from the ideal Krylov states $\hat{U}_{m}\ket{r}$; \emph{given} those prepared states, however, both $S_{m'r',mr}=\braketmid{r'}{\hat{V}_{m'}^{\dagger}\hat{V}_{m}}{r}$ and $H^{\mathrm{NT}}_{m'r',mr}=\braketmid{r'}{\hat{V}_{m'}^{\dagger}\hat{H}\hat{V}_{m}}{r}$ are exact, so the entire Trotterization error is absorbed into the basis states rather than into the matrix entries. The pair $(H^{\mathrm{NT}},S)$ is consequently a mutually consistent (Galerkin) representation of $\hat{H}$ on the prepared subspace, departing from the ideal-basis representation only through the state drift $\hat{V}_{m}-\hat{U}_{m}$ (finite sampling and real noise sources aside). The Toeplitz approximation, by contrast, perturbs $H$ alone, introducing an inconsistency with $S$ bounded by equation~\ref{ec:toeplitz_bound}; this anticipates the behavior reported in Section~\ref{sec:results}, where the non-Toeplitz formulation is found to be weakly sensitive to $\varepsilon_{T}$ while the Toeplitz formulation tracks it.
Second, the bound is conservative: the use of $\|\hat{H}\|_{1}\geq\|\hat{H}\|$ and the worst-case index $m=M-1$ both loosen it, and it constrains the matrix entries, with a spectral-norm statement on the full $MR\times MR$ matrix incurring an additional dimensional factor.

Despite this additional error, we retain the Toeplitz approximation as a resource-reduction heuristic. It reduces the number of distinct matrix elements requiring estimation to $\approx MR + \frac{1}{2}(2M-1)(R-1)R$ and, because each diagonal is populated by a single approximate propagator rather than two independently approximated ones, roughly halves the per-element circuit depth and gate count while limiting the accumulation of product-formula overhead across $S$ and $H$. Under this approximation the matrix elements take the form
\begin{eqnarray}
    \label{ec:S_and_H_elements}
    S_{m'r',\:mr} &=& \braketmid{r'}{e^{-i\p{m-m'}\tau\hat{H}}}{r}, \\ 
    H_{m'r',\:mr} &=& \braketmid{r'}{\hat{H}e^{-i\p{m-m'}\tau\hat{H}}}{r},
\end{eqnarray}
which, in our numerical simulations, are evaluated using Hadamard tests with the second-order ($k=2$) product formula introduced above.

For all three configurations, we evaluated the autocorrelation function $C(t) = |\braket{\psi_{0}}{\psi(t)}|^{2}$ with $R=5,10,15$, taking as reference states the bitstrings of largest measurement probability in $e^{-i(M-1)\tau\hat{H}}\ket{\psi_{0}}$, following the procedure of Ref.~\cite{sQKFF_Algorithm}.

To isolate the influence of reference-state selection, we compared the autocorrelation function of Configuration~1 ($R=10$, $\varepsilon_{T}=0.09392$) across six construction strategies: (i) highest-probability bitstrings sampled at $T=(M-1)\tau$; (ii) highest-probability bitstrings sampled at $T'=2(M-1)\tau$; (iii) randomly selected bitstrings; (iv) randomly selected product states; and (v)--(vi) reference states drawn from QWC-group measurements. In the last approach, letting $\mathcal{P}_{i}$ denote the Clifford rotation that diagonalizes the qubit-wise commuting group $\mathcal{G}_{i}$, we measured both $\mathcal{P}_{i}^{\dagger}\ket{\psi_{0}}$ and $\mathcal{P}_{i}^{\dagger}e^{-i(M-1)\tau\hat{H}}\ket{\psi_{0}}$, mapped the highest-probability outcomes $\ket{k_i}$ back to the computational basis as $\mathcal{P}_{i}\ket{k_i}$, and added them to the reference-state pool. These states were expected to combine sizable overlap with the dynamically relevant subspace with straightforward preparation. Since one of the commuting groups is purely diagonal, the second measurement scheme reproduces the original reference-selection procedure of Ref.~\cite{sQKFF_Algorithm}, 

Finally, to assess the Toeplitz approximation in practice, the autocorrelation function of Configuration~1 was also computed without imposing the Toeplitz structure on $S$ and $H$, for $\varepsilon_{T}=0.41858$ and $0.09392$, enabling a direct comparison with the Toeplitz-approximated results. The results of these experiments are discussed in the forthcoming.

\subsubsection{Scaling of the method.}

A central question for any fast-forwarding method (especially subspace methods) in this context is whether it remains viable as the number of interacting defects $N$ grows, given that the dimension of the underlying Hilbert space grows exponentially in $N$. It is useful to separate two distinct cost channels: the cost of preparing and measuring each quantum circuit, and the cost of spanning the dynamically relevant subspace with reference states. We argue that the former scales benignly in $N$, and that the exponential growth of the Hilbert space enters the problem (if at all) only through the latter.

Every per-circuit and per-element cost of the algorithm is polynomial in $N$. Under the Gray encoding, the qubit count grows linearly, each defect contributing $\lceil\log_{2}d_{\mathrm{electron}}\rceil + \lceil\log_{2}d_{\mathrm{nuclear}}\rceil$ qubits together with a shared bath register of $\lceil\log_{2}d_{\mathrm{bath}}\rceil$ qubits. The number of Pauli terms in $\hat{H}$ grows as $\mathcal{O}(N^{2})$, dominated by the pairwise dipole--dipole network and built on the $\mathcal{O}(N)$ local zero-field, Zeeman, hyperfine, and spin--boson terms. The depth of a single Trotter step is therefore $\mathcal{O}(N^{2})$ before optimization and sub-quadratic once qubit-wise commuting aggregation is applied \cite{Verteletskyi2019, JoshIzaac2021}. Under the Toeplitz structure of Sec.~II.C, the number of distinct Krylov matrix elements to be estimated is $\mathcal{O}(MR^{2})$ and is independent of $N$ at fixed subspace dimension. Estimating each element to additive precision $\varepsilon$ requires $\mathcal{O}(\|\hat{H}\|_{1}^{2}/\varepsilon^{2})$ samples; since $\|\hat{H}\|_{1}=\mathcal{O}(N)$ for the Hamiltonians considered here, this amounts to $\mathcal{O}(N^{2}/\varepsilon^{2})$ shots per element, to which the shot-reduction strategies discussed in Sec.~\ref{sec:consid_and_outlook} further apply. None of these costs is prohibitive.

The exponential dimension of the Hilbert space thus does not enter through the per-circuit cost. It enters, if at all, through a single channel: the number of reference states $R$ needed to span the dynamically relevant subspace, and hence the Krylov dimension $MR$. The statement that $MR$ is independent of $N$ holds only at fixed $R$, and holding $R$ fixed is not free. The decisive quantity is how the support of the evolving state is distributed over the eigenbasis of $\hat{H}$. When reference states with substantial overlap onto the states traversed during the dynamics can be identified, a modest, slowly growing $R$ suffices and the method scales favorably with system size. When such states cannot be found, the no-fast-forwarding theorems \cite{Atia2017} are consistent with the worst case in which the Krylov dimension required for a fixed target accuracy grows exponentially with $N$. Reference-state quality is, in this sense, the binding constraint on scalability, rather than circuit depth, gate count, or measurement overhead.

This expectation is already visible in our numerical results. Configuration~2, which has the largest Hilbert space among the cases studied, distributes its measurement probability mass across many bitstrings, so that at fixed $R$ fewer of the dynamically important components are retained and the long-time reconstruction degrades accordingly (Sec.~\ref{sec:results}). The same mechanism that controls accuracy at the system sizes accessible here is therefore expected to control scalability at larger ones. This identifies dynamics-aware reference-state selection as the principal lever for extending the framework to large interacting ensembles, and motivates the adaptive and residual-based selection strategies discussed in Sec.~\ref{sec:consid_and_outlook}.

\subsection{Microwave Absorption Spectra Computation via sQKFF}
Most spin-defect-based platforms exhibit optical activity when interacting with microwave magnetic fields. This property forms the foundation of the Optical Detection of Magnetic Resonance (ODMR) technique, which has been crucial for precise measurements and high-fidelity control of the spin quantum state. Spin defects are promising candidates for quantum technologies due to the capacity to manipulate the spin state reliably, e.g., via coupling to the magnetic component of incident electromagnetic waves, combined with coherence times long enough to apply these interactions effectively \cite{Fang2024}.

Research focusing on the applications and operational enhancement of spin defects through microwave and optical frequency interactions is highly active. The resulting applications span a wide range of quantum technologies, including quantum memories, scalable quantum computing and sensing, as well as enabling connections to photonic architectures for the implementation of quantum networks \cite{Morton2018, Cache2025, Resch2025, Chowdhury2024}. Therefore, providing a way to calculate the microwave absorption spectra would demonstrate the operational utility of simulating these spin-defect-based platforms, advancing the study and development of these systems for quantum technologies.

Linear response applies when the perturbation is sufficiently weak, and the system is initially in a stationary state of $\hat{H}_{0}$ (typically thermal equilibrium). We assume a static bias field defining the $\hat{z}$ quantization axis and probe the system with a weak transverse oscillating magnetic field,
\begin{equation}
\hat{H}_{P}(t) = -B_{x}\cos(\omega t)\, \hat{M}_{x}.
\end{equation}
In the long-wavelength limit  $\lambda \gg L$, where $L$ is the length scale of the system of interest, the driving field is effectively uniform across the spin ensemble; this is the magnetic dipole approximation, which we adopt in the forthcoming. For NV$^{-}$ center configurations, $\hat{M}_{x}$ is the $x$-component of the total magnetic moment, including the electronic spin operator $\hat{S}_{x}$ and nuclear spin operators $\hat{I}_{x}^{(j)}$, with each contribution weighted by its gyromagnetic ratio.

The linear absorption spectrum follows from the imaginary part of the susceptibility $\chi(\omega)$, $\chi''(\omega)$, obtained from the Fourier transform of the retarded correlation function $\mathcal{C}(t) = -i\braketmid{\psi_0}{[e^{it\hat{H}}\hat{M}_{x}e^{-it\hat{H}}, \hat{M}_{x}]}{\psi_{0}}$, with $\hat{H}$ the stationary system's Hamiltonian. The absorbed power scales as $\omega\:\chi''(\omega)$. Defining $Z(t) = \braketmid{\psi_{0}}{e^{it\hat{H}}\hat{M}_{x}e^{-it\hat{H}}\hat{M}_{x}}{\psi_{0}}$, it follows $\mathcal{C}(t)=2\:Im(Z(t))$. The sQKFF algorithm is employed to compute $Z(t)$ by approximating the time-evolved states $\ket{\psi(t)}$ and $\ket{\eta(t)}=e^{-it\hat{H}}\hat{M}_x\ket{\psi_0}$ in their respective Krylov subspaces, with independently sampled reference bitstrings. This yields

\begin{equation}
    \label{ec:Z_fnc_approximation}
    Z(t)\approx (\textbf{c}^{\psi}(t))^{\dagger}A\textbf{c}^{\eta}(t),
\end{equation}

\noindent where the $\textbf{c}(t)$ are obtained from the Krylov-projected Schrödinger equation, \ref{ec:Krylov_Schrodinger_eq}, and $A$ is an $MR\times MR'$ matrix with elements $A_{(mr), (m'r')} = \braketmid{\phi_{mr}^{\psi}}{\hat{M}_{x}}{\phi_{m'r'}^{\eta}}$. These matrix elements are computed using Hadamard tests combined with an LCU decomposition for the non-unitary operator $\hat{M}_x$.

\subsection{$\ell_{1}$-norm of coherence with sQKFF}
Assessing the coherence time of a quantum system typically requires comparing the coherence of its time-evolved state against that of the initial state. A plethora of coherence measures have been proposed in the literature to quantify this basis-dependent resource \cite{RevModPhys.89.041003, Aberg2006Dec, PhysRevLett.113.140401, PhysRevA.92.022124,Albrecht2013, PhysRevLett.116.120404, PhysRevA.94.052336, deVicente2016Dec, PhysRevLett.122.190405,PhysRevResearch.2.023298}. Among these, one of the widely employed resource quantifiers is the $\ell_1$-norm of coherence \cite{PhysRevLett.113.140401} due to its operational clarity.

Expressing the state of an $n$-qubit quantum system, $\hat{\rho}$, in the computational basis $\Theta\equiv \{\ket{i}\:|\: i\in\{0,1\}^{n}\}$, the $\ell_1$-norm of coherence, denoted $C_{\ell_1}(\hat{\rho})$, is given by
\begin{equation}
    \label{ec:l1_norm_of_coherence}
    C_{\ell_1}(\hat{\rho}) = \sum_{\{i\neq j \: | \: i,j \in \Theta\}}|\bra{i}\hat{\rho}\ket{j}|\,.
\end{equation}

From the above definition, it is evident that measuring coherence is experimentally costly. For a density matrix representing the state of the system, computing $2^{2n} - 2^n$ off-diagonal parameters is required in the worst-case scenario. Moreover, when studying dynamics, the cost escalates further, as coherence must be evaluated at every time point. However, when the system evolves within a Krylov subspace, this overhead can be significantly reduced. 

The system studied in this work is a number $\mathcal{N}$ of NV$^{-}$ centers coupled with an environment, which includes the nuclear spins of each NV$^{-}$ center, the spin of $^{14}$N impurities, and a phonon bath. Considering it as a bipartite system composed of subsystems $A$, the NV$^{-}$ electronic spins, and $B$, their environment, the time-evolved state of the complete system is
\begin{equation}
    \label{ec:coherence_time_ev_state}
    \ket{\psi}_{AB}\equiv\ket{\psi(t)} = e^{-it\hat{H}/\hbar}\ket{\psi_{0}},
\end{equation}

\noindent which is approximated by equation \ref{ec:krylov_approx}. Then, in the Krylov approximation, the $\ell_1$-norm as defined in equation \ref{ec:l1_norm_of_coherence} can be expressed as
\begin{equation}
    \label{ec:l1_coherence_krylov}
    \begin{split}
        & C_{\ell_{1}}\p{\ket{\psi}\bra{\psi}_{AB}} \\
        & \quad = \sum_{\substack{k,l\in\Sigma\times\Gamma \\ k\neq l}}\left|\sum_{m,r}\sum_{m',r'}c_{mr}(t)c_{m'r'}^{*}(t)\braket{k}{\phi_{mr}}\braket{\phi_{m'r'}}{l}\right|,
    \end{split}
\end{equation}

\noindent with binary alphabets $\Sigma$ and $\Gamma$ for subsystems $A$ and $B$, respectively, such that their computational basis are $\{\ket{a}\:|\: a\in\Sigma\}$ and $\{\ket{b}\:|\: b\in\Gamma\}$.

Now, computing the coherence of the state at any time is only dependent on classically obtaining $\mathbf{c}(t)$ from equation \ref{ec:Krylov_Schrodinger_eq}, which doesn't require an additional quantum effort provided we already solved the Krylov subspace problem. The amplitudes of the corresponding computational basis states that compose the reference states $\ket{\phi_{mr}}$ can always be obtained via Hadamard tests, and remain the same at all times $t$. However, since it is usual for reference states to be classically tractable,  and we require only inner products of these states with computational basis bitstrings, the extra computational efforts required to compute the $\ell_1$ norm are typically classical. Thus, while the process of computing the amplitudes may scale exponentially with the number of qubits, all subsequent coherence evaluations are efficient.

\begin{figure*}
    \centering
    \begin{subfigure}[b]{0.49\textwidth}
        \includegraphics[width=0.9\linewidth]{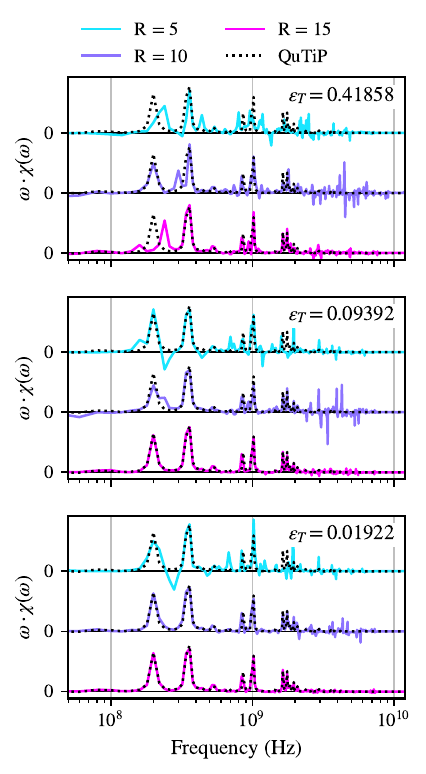}
        \caption{Microwave absorption spectra, normalized to the peak intensity of the QuTiP result.}
        \label{fig:Spectrum_comparison}
    \end{subfigure}
    \hfill
    \begin{subfigure}[b]{0.49\textwidth}
        \includegraphics[width=0.9\linewidth]{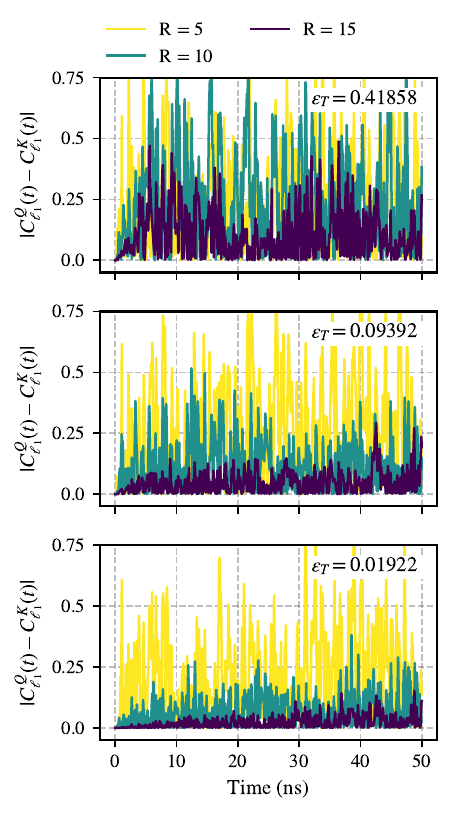}
        \caption{Absolute difference $|C_{\ell_{1}}^{Q} - C_{\ell_{1}}^{K}|$ as a function of time.}
        \label{fig:coherence}
    \end{subfigure}
\caption{Benchmark of the framework against exact (QuTiP) results for Configuration 1, showing that sQKFF reproduces both observables and that accuracy improves as the number of reference states $R$ increases and the Trotterization error $\varepsilon_{T}$ decreases. (a) The sQKFF microwave absorption spectrum recovers the dominant resonance peaks of the QuTiP reference, resolving all significant peaks at the smallest $\varepsilon_{T}$ and largest $R$. (b) Absolute difference between the $\ell_{1}$-norm of coherence from QuTiP, $C_{\ell_{1}}^{Q}$, and from sQKFF, $C_{\ell_{1}}^{K}$, versus time; the error grows more slowly for larger $R$, indicating that the reference-state count is the dominant control on long-time coherence accuracy.}
\label{fig:spectrum_and_coherence}
\end{figure*}

Tracing out the environment and rearranging terms yields a simplified expression for computing the $\ell_{1}$ norm of coherence of our system of interest: the NV$^{-}$ electronic spins, given by
\begin{equation}
    \label{ec:l1_coherence_subsystem}
    C_{\ell_{1}}\p{\hat{\rho}_{A}} = \sum_{\{i\neq j \:|\: i,j\in\Sigma\}}\sum_{\{b\:|\:b\in\Gamma\}}\left|\gamma_{ib}\gamma_{jb}^{*}\right|
\end{equation}

\noindent with $\gamma_{ib}=\sum_{mr}c_{mr}(t)\braket{ib}{\phi_{mr}}$, where $\ket{ib}=\ket{i}\otimes\ket{b}$. Hence, the coherence of the subsystem’s evolved state can be computed efficiently once the reference states' amplitudes are known. As a direct consequence, within the Krylov subspace, it is efficient to determine the time $t$ when the coherence of the system decays to $1/e$ of its initial value, thus providing valuable information related to effective $T_1$ and $T_2$ coherence times.

\section{Results \label{sec:results}}
Figure \ref{fig:Spectrum_comparison} compares the microwave absorption spectrum of Configuration 1 with QuTiP approximations. Even at relatively large Trotterization errors, the framework correctly identifies the dominant resonance frequencies under microwave driving. Accuracy systematically improves as the Trotterization error $\varepsilon_{T}$ decreases and the number of reference states $R$ increases, ultimately resolving all significant peaks. This agreement confirms that the proposed approach reliably captures operational properties derived from system dynamics, supporting the study, modeling, and advancement of spin defect-based platforms.

The $\ell_{1}$-norm of coherence for Configuration 1 in figure \ref{fig:coherence} exhibits trends consistent with the spectral results: decreasing $\varepsilon_{T}$ enhances accuracy, while increasing $R$ enables reliable simulations over operationally relevant timescales without excessive resource growth. The stronger sensitivity to $R$ highlights the importance of reference-state selection. 

\begin{figure*}
    \centering
    \includegraphics[width=0.8\linewidth]{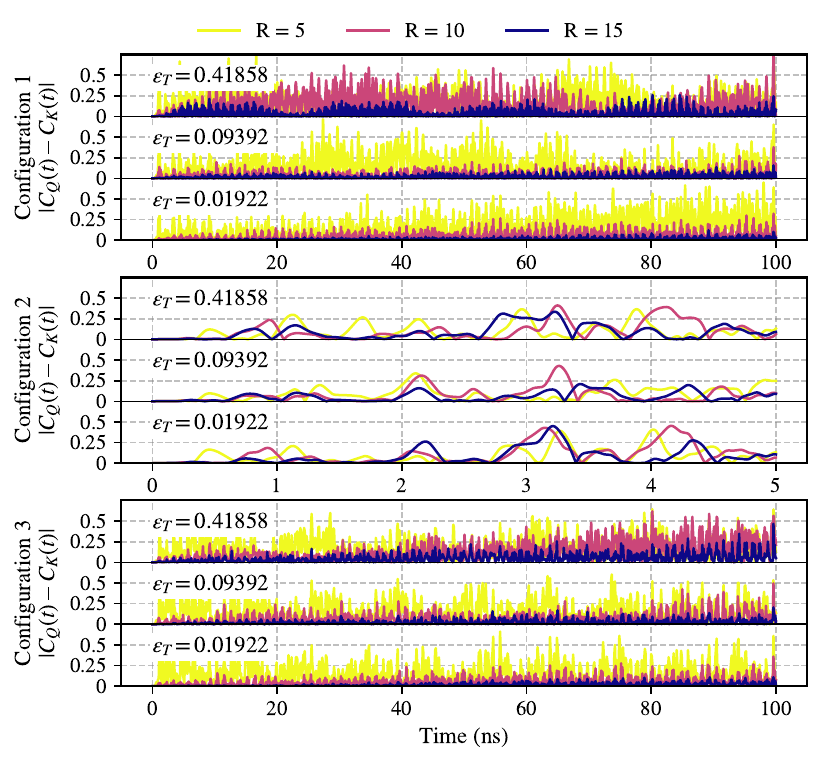}
    \caption{Autocorrelation-function benchmark across all three configurations, showing that agreement with the exact (QuTiP) result extends to longer times as the number of reference states $R$ increases, and that this extension (as opposed to the Trotterization error $\varepsilon_{T}$) is the dominant factor. Plotted is the absolute difference $|C_{Q}(t)-C_{K}(t)|$ between the QuTiP and sQKFF autocorrelation functions for each $(\varepsilon_{T},R)$ combination. Configurations 1 and 3 remain accurate for several tens of nanoseconds, whereas Configuration 2, with the largest Hilbert space, is reproduced less accurately at fixed $R$ because its measurement probability is spread across many bitstrings.}
    \label{fig:autocorrelation}
\end{figure*}

Figure \ref{fig:autocorrelation} further illustrates the interplay between $\varepsilon_{T}$ and $R$ in computing autocorrelation functions. For Configurations 1 and 3, the sQKFF results agree with QuTiP for several tens of nanoseconds. Notably, short-time dynamics ($\approx 20$ ns) are reproduced with reasonable precision even at larger $\varepsilon_{T}$, implying that shallower circuits can still yield meaningful predictions. However, increasing $R$ substantially extends both accuracy and accessible simulation time. 

Configuration 3 is particularly illustrative: despite its larger Hilbert space and additional dipole-dipole and impurity interactions, the case $\varepsilon_{T}=0.01922$, $R=15$ achieves accuracy comparable to Configuration 1. This suggests that reference-state selection seems to dominate the long-time error budget, over either Trotterization error or system size within the tested parameters regime and the representative cases studied. These observations motivate the development of adaptive or physically informed reference-state strategies that may improve long-time accuracy while reducing quantum resource requirements.

In contrast, Configuration 2 shows reduced performance in figure \ref{fig:autocorrelation}. Its larger Hilbert space distributes measurement probabilities across many bitstrings, complicating the identification of reference states that capture extended-time features. Additionally, its larger Hamiltonian 1-norm, $\|\hat{H}\|_{1}/\hbar = 112.265$ GHz compared to $62.1052$ and $62.1124$ GHz for Configurations 1 and 3, respectively, affects $\tau$ and thus the construction of reference states. This suggests that an optimal sampling time $T=(M-1)\tau$ may exist at which measured bitstrings encode maximal information about accessible-time dynamics, warranting systematic investigation of the interplay between $\|\hat{H}\|_{1}$, $R$, and $\varepsilon_{T}$.

\begin{figure*}
    \centering
    \includegraphics[width=0.85\linewidth]{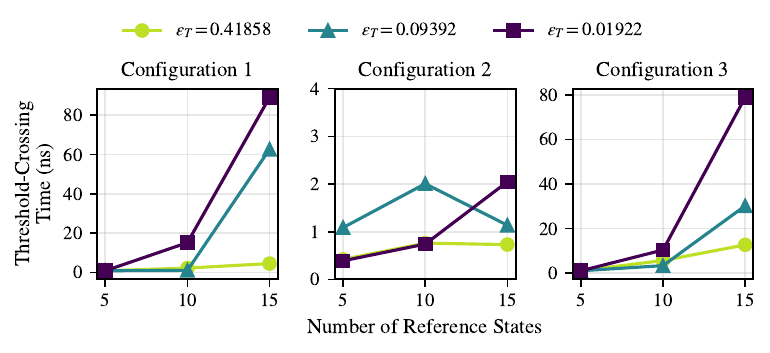}
    \caption{Threshold-crossing time of the absolute difference between the autocorrelation function approximated via QuTiP and sQKFF, defined as the earliest time $t$ at which $|C_{Q}(t) - C_{K}(t)|>0.1$, shown as a function of the number of reference states for the three Trotterization errors considered in this work.}
    \label{fig:threshold_crossing}
\end{figure*}

In the idealized limit of exact time evolution and a finite-dimensional invariant subspace fully captured by the selected reference states, the sQKFF can, in principle, reconstruct dynamics at arbitrarily long times. In practice, however, the accessible evolution time depends strongly on the quality and number of reference states, the Trotterization error, and the extent to which the dynamically relevant subspace is represented, with the accessible-time improvement expected to eventually saturate once the relevant subspace is sufficiently spanned. Figure \ref{fig:threshold_crossing} provides an empirical characterization of this behavior by showing the threshold-crossing time, defined as the earliest time at which $|C_{Q}(t)-C_{K}(t)|>0.1$, as a function of the number of reference states. The results indicate a consistent increase in accessible evolution time as $R$ grows for the tested Trotterization errors. At present, no general analytical scaling law relating the accessible evolution time to $R$ is known beyond the convergence considerations discussed in Refs. \cite{sQKFF_Algorithm, sQK_Algorithm}.

\begin{table*}[!t]
    \caption{Per-circuit resource estimates for each Hadamard-test circuit used to compute the matrix elements in equation \ref{ec:S_and_H_elements}, shown for all configurations and Trotterization thresholds $\varepsilon_{T}$. Second-order Trotterization was employed, with $\varepsilon_{T}=0.41858$, $0.09392$, and $0.01922$ corresponding to 9, 19, and 42 steps, respectively. $\mathcal{G}\:[\times10^{4}]$ and $\mathcal{W}\:[\times10^{4}]$ denote resource counts with and without QWC aggregation, and $\gamma\:(\%)$ is the relative reduction. $T$-gate counts per $Rz$ rotation assume precision $\varepsilon = 10^{-6}$ \cite{qualtran}.
    }
    \centering
    {
    \begin{tabular}{r|ccccccccc}
    \hline
    \multicolumn{1}{c|}{$\varepsilon_{T}$}& \multicolumn{3}{c|}{0.41858}                                                  & \multicolumn{3}{c|}{0.09392}                                                  & \multicolumn{3}{c}{0.01922}                              \\ \hline
    \multicolumn{1}{c|}{\textbf{Metric}} & \multicolumn{1}{c|}{\textbf{$\mathcal{W}$}} & \multicolumn{1}{c|}{\textbf{$\mathcal{G}$}} & \multicolumn{1}{c|}{\textbf{$\gamma$}}      & \multicolumn{1}{c|}{\textbf{$\mathcal{W}$}} & \multicolumn{1}{c|}{\textbf{$\mathcal{G}$}} & \multicolumn{1}{c|}{\textbf{$\gamma$}}      & \multicolumn{1}{c|}{\textbf{$\mathcal{W}$}} & \multicolumn{1}{c|}{\textbf{$\mathcal{G}$}} & \textbf{$\gamma$}      \\ \hline
    \multicolumn{1}{l|}{}       & \multicolumn{9}{c}{\textbf{Configuration 1}}                                                                                                                                                                                      \\ \hline
    \textbf{H}                    & 1.6                    & 1.2                    & \multicolumn{1}{c|}{27.17} & 3.4                    & 2.5                    & \multicolumn{1}{c|}{27.2} & 7.5                    & 5.5                    & 27.22 \\
    \textbf{CNOT}                        & 4.4                    & 3.1                    & \multicolumn{1}{c|}{29.94} & 9.3                    & 6.5                    & \multicolumn{1}{c|}{29.94} & 20.5                   & 14.4                   & 29.95 \\
    \textbf{T}                           & 82.3                   & 64.3                   & \multicolumn{1}{c|}{21.87} & 173.6                  & 135.6                  & \multicolumn{1}{c|}{21.89} & 383.7                  & 299.6                  & 21.9 \\
    \textbf{Depth}               & 7.5                    & 5.3                    & \multicolumn{1}{c|}{29.16} & 15.9                   & 11.3                   & \multicolumn{1}{c|}{29.17} & 35.1                   & 24.9                   & 29.17 \\ \hline
    \multicolumn{1}{l|}{}       & \multicolumn{9}{c}{\textbf{Configuration 2}}                                                                                                                                                                                      \\ \hline
    \textbf{H}                    & 6.4                    & 4.7                    & \multicolumn{1}{c|}{26.84}                      & 13.5                   & 9.9                    & \multicolumn{1}{c|}{26.85}                      & 29.8                   & 21.8                   & 26.85 \\
    \textbf{CNOT}                        & 18.5                   & 13.4                   & \multicolumn{1}{c|}{27.8}                      & 39.1                   & 28.2                   & \multicolumn{1}{c|}{27.8}                      & 86.5                   & 62.4                   & 27.81 \\
    \textbf{T}                           & 320.6                  & 250.2                  & \multicolumn{1}{c|}{21.96}                      & 676.6                  & 528.0                  & \multicolumn{1}{c|}{21.97}                      & 1 495.4                 & 1 166.9                 & 21.97 \\
    \textbf{Depth}               & 31.4                   & 22.8                   & \multicolumn{1}{c|}{27.34}                      & 66.3                   & 48.2                   & \multicolumn{1}{c|}{27.34}                      & 146.5                  & 106.5                  & 27.35 \\ \hline
    \multicolumn{1}{l|}{}       & \multicolumn{9}{c}{\textbf{Configuration 3}}                                                                                                                                                                                      \\ \hline
    \textbf{H}                    & 3.1                    & 2.3                    & \multicolumn{1}{c|}{23.62} & 6.5                    & 4.9                    & \multicolumn{1}{c|}{23.63} & 14.3                   & 10.9                   & 23.64 \\
    \textbf{CNOT}                        & 8.5                    & 6.3                    & \multicolumn{1}{c|}{25.65} & 17.9                   & 13.3                   & \multicolumn{1}{c|}{25.65} & 39.5                   & 29.4                   & 25.65 \\
    \textbf{T}                           & 156.5                  & 126.9                  & \multicolumn{1}{c|}{18.95} & 330.3                  & 267.7                  & \multicolumn{1}{c|}{18.96} & 730.0                  & 591.6                  & 18.97 \\
    \textbf{Depth}               & 14.4                   & 10.8                   & \multicolumn{1}{c|}{25.04} & 30.5                   & 22.9                   & \multicolumn{1}{c|}{25.04} & 67.4                   & 50.5                   & 25.05 \\ \hline
    \end{tabular}
    }
    \label{tab:ResourceEstimations}
\end{table*}

Resource estimates for constructing the $S$ and $H$ matrices in table \ref{tab:ResourceEstimations} confirm substantial reductions in per-circuit gate count and circuit depth from QWC aggregation, ranging between $18-30\%$. These reductions are particularly relevant for NISQ implementations, where the feasibility of individual Hadamard-test evaluations is strongly limited by coherent error accumulation and decoherence. The reduction ratio $\gamma$ remains approximately constant as $\varepsilon_{T}$ decreases, consistent with the repeated structure of the Trotterized evolution, with minor variations arising from adjoint cancellations between successive steps.

\begin{figure*}
    \centering
    \includegraphics[width=0.55\linewidth]{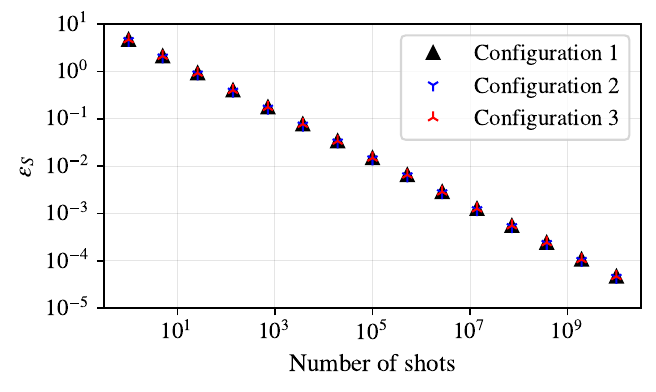}
    \caption{Relative error estimation, $\varepsilon_{S}$, in computing the expectation value of $\hat{H}$. Results are shown for all configurations and their initial states described in table \ref{tab:TableParameters}.}
    \label{fig:shot_error}
\end{figure*}

Figure \ref{fig:shot_error} shows the estimated statistical error $\varepsilon_{S}$ as a function of the number of measurement shots for the three configurations considered in this work. In all cases, the error decreases approximately as $N_{shots}^{1/2}$, consistent with the expected scaling of sampling uncertainty in measurements \cite{Gonthier2022}. The three configurations exhibit nearly identical behavior, indicating the shot cost is primarily determined by the target precision rather than details of the underlying spin-defect system. Then, achieving an improvement of one order of magnitude in precision requires approximately two additional orders of magnitude in the number of shots. These results highlight that, although the proposed framework reduces circuit depth and gate count, measurement cost remains a significant practical consideration, as the shot budget grows quadratically with the precision. 

\subsection{On the comparison of the reference-state selection strategies.}

\begin{figure*}
    \centering
    \includegraphics[width=0.9\linewidth]{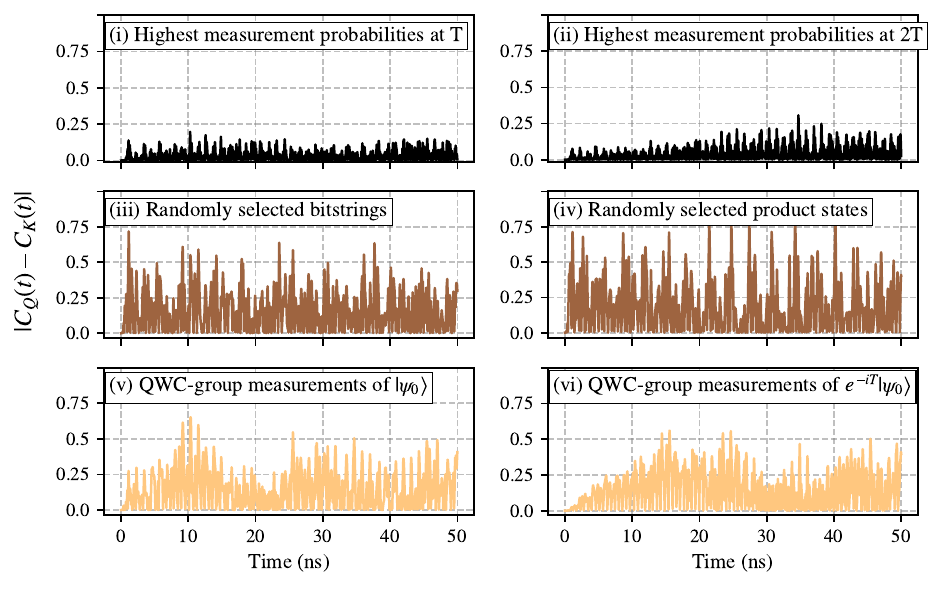}
    \caption{Time evolution of the absolute difference between the autocorrelation function computed with QuTiP, $C_{Q}(t)$, and those obtained with sQKFF, $C_{K}(t)$, for Configuration 1 with $R=10$ and $\varepsilon_{T} = 0.09392$. Each graph corresponds to a different reference-state selection strategy as labeled.}
    \label{fig:autocorrelation_ref_states}
\end{figure*}

Figure \ref{fig:autocorrelation_ref_states} illustrates the influence of the reference-state selection strategy on the quality of the Krylov reconstruction. Strategies based on selecting bitstrings with the highest measurement probabilities provide the most accurate time approximations among the tested approaches, whereas randomly selected bitstrings and random product states exhibit substantially poorer performance as the evolution time increases.

Although the QWC-group measurement schemes perform less accurately than the highest-probability bitstring selection, they still retain acceptable agreement with the QuTiP approximation over a limited time window (below approximately 5 ns for scheme (vi)). These results support the physically motivated heuristic underlying the reference-state selection procedure, namely that states exhibiting larger overlap with the time-evolved initial state bias the Krylov subspace toward the dynamically relevant sector of Hilbert space.

Nevertheless, as discussed in Sec.~II, the present strategy does not provide a provably optimal reference-state selection, and the accuracy of the long-time reconstruction remains sensitive to the quality and number of selected reference states.

\subsection{On the effects of the Toeplitz structure for Krylov subspace matrices.}

\begin{figure*}
    \centering
    \includegraphics[width=0.5\linewidth]{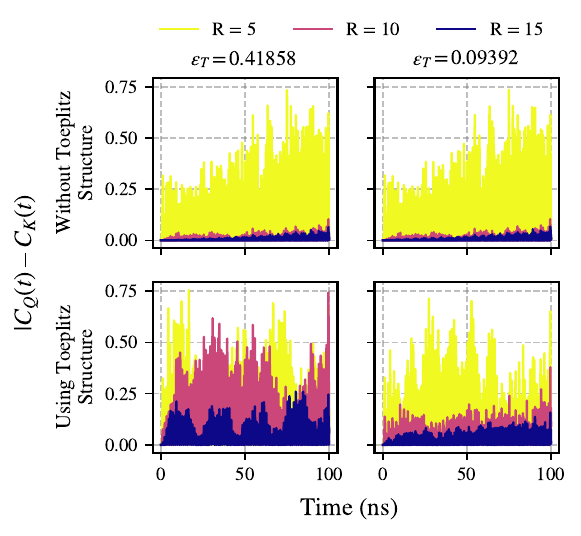}
    \caption{Absolute difference between the autocorrelation functions computed with QuTiP, $C_{Q}(t)$, and sQKFF, $C_{K}(t)$, for Configuration 1, comparing the Toeplitz and non-Toeplitz formulations of the Krylov matrices S and H. Results are shown for two Trotterization error values $\varepsilon_{T}$, illustrating the impact of the Toeplitz approximation on the accuracy of the reconstructed dynamics.}
    \label{fig:toeplitz_comparison}
\end{figure*}

Figure \ref{fig:toeplitz_comparison} illustrates the effect of imposing the Toeplitz structure on the Krylov matrices $S$ and $H$ for Configuration 1. The non-Toeplitz formulation exhibits only a weak dependence on the Trotterization error, whereas the Toeplitz-structured results follow the error scaling discussed in Section II.C. In particular, the discrepancy between the Toeplitz and non-Toeplitz approximations increases with the Trotterization error, consistent with the bound derived for the Toeplitz approximation.

Despite this loss in accuracy, both formulations benefit substantially from increasing the number of reference states, reassuring its significance in the long-time approximation accuracy. The non-Toeplitz approach generally provides a more accurate reconstruction of the autocorrelation function, but at a significantly higher quantum-resource cost. Constructing the Krylov Hamiltonian without the Toeplitz approximation requires independent estimation of matrix elements containing two product-formula evolutions, resulting in an approximate doubling of the gate count and circuit depth relative to the Toeplitz case. Moreover, the number of matrix elements that must be estimated scales as $\mathcal{O}(M^{2}R^{2})$, compared with $\mathcal{O}(MR^{2})$ when the Toeplitz structure is assumed.

These results highlight a trade-off between accuracy and resource requirements. While the Toeplitz approximation introduces an additional source of error that becomes increasingly relevant at longer evolution times and larger Trotterization errors, it substantially reduces circuit depth, gate count, and measurement overhead. Consequently, the suitability of the approximation depends on the available hardware resources, target accuracy, and the extent to which error-mitigation or error-correction techniques can compensate for the additional approximation error. Within the context of NISQ devices, the reduced circuit complexity afforded by the Toeplitz structure may outweigh the associated loss in accuracy.

\section{Additional Considerations and Outlook \label{sec:consid_and_outlook}}
The framework presented in this work shows that simulating the dynamics of a quantum system over short timescales can enable access to physically and operationally relevant regimes. In particular, our numerical results using a few picoseconds of quantum simulation show excellent agreement with exact simulations up to 50–100 nanoseconds. This proof of principle indicates that, even under restrained resources and hardware noise, quantum computers combined with optimized simulation techniques can serve as effective platforms for studying the properties and dynamics of spin defects in solid-state materials.

The motivation for developing such approaches is reinforced by the intrinsic computational complexity of quantum many-body dynamics. Recent quantum-computing demonstrations have reached dynamical regimes that are believed to be classically intractable. For example, large-scale simulations of spin hamiltonians have provided evidence of quantum advantage in the study of non-equilibrium spin dynamics \cite{King2025}, while more recent experiments involving higher-order out-of-time-ordered correlators (OTOCs) have shown that certain dynamical observables cannot be accurately reproduced even by current classical supercomputing resources \cite{Google2025}. These results highlight the growing gap between the capabilities of classical algorithms and the complexity of experimentally relevant quantum dynamics.

While the representative NV systems used to benchmark the framework remain classically tractable, the methodology is intended for regimes in which conventional classical approaches become increasingly challenging to scale. For example, cluster-correlation expansion methods may lose accuracy in dense interacting-defect ensembles with strong dipole-dipole coupling networks, while tensor-network approaches such as matrix-product-state (MPS) methods can become limited by rapid entanglement growth, particularly in non-perturbative spin-boson regimes near the polaronic crossover. In addition, exact-diagonalization-based approaches scale exponentially with system size, restricting their applicability to relatively small Hilbert spaces. Consequently, quantum-computing-based methodologies may provide a viable pathway for investigating spin-defect dynamics in parameter regimes that are inaccessible to existing classical techniques.

Advancing toward practical implementation on real quantum hardware will require addressing additional considerations, such as the number of measurement shots, crucial for accurately estimating expectation values \cite{Crawford2021, Yen2021}, as well as the number of qubits and their limited connectivity \cite{Brierley2017, Holmes2020}. The former is relevant because the sQKFF algorithm relies on computing the expectation value of the Hamiltonian, whereas the latter becomes particularly relevant when simulating quantum systems with non-trivial interactions, such as the dipole–dipole coupling present in Configurations 2 and 3, which introduces a large number of entangled pairs. 

Although the QWC decomposition substantially reduces per-circuit gate count and circuit depth, measurement overhead remains a significant cost. Estimating each matrix element with additive precision $\epsilon$ through direct sampling generally requires $\mathcal{O}(1/\epsilon^{2})$ repetitions, and finite-shot fluctuations can be amplified by the conditioning of the Krylov overlap matrix $S$. Nevertheless, the Toeplitz approximation reduces the number of distinct matrix elements that must be estimated, while the QWC partitioning also enables grouped measurements of commuting Pauli operators, independently lowering the total shot budget. Under realistic noisy-device conditions, the reduction in circuit depth becomes increasingly significant, since shallower circuits accumulate less coherent control error and decoherence.

Krylov-subspace methods are known to be particularly sensitive to noise because the overlap matrix $S$ may become ill-conditioned, so practical implementations often regularize the problem by projecting onto subspaces associated with eigenvalues above a chosen threshold \cite{Kirby2024, Yoshioka2025}. In sQKFF, statistical fluctuations originating from finite-shot Hadamard tests affect both the estimated Krylov matrices and the probability distributions used for reference-state selection. Consequently, wall-clock execution time may become dominated by sampling cost when circuit evaluations are serialized. Several complementary mitigation strategies are available, including recent approaches for reducing shot overhead through symmetry-adapted subspace methods \cite{Lee2025, patel2025quantumsenioritybasedsubspaceexpansion}, distributed shot-parallel execution across multiple QPUs \cite{Bisicchia2024ShotWise, Barral2025ReviewDQC}, and amplitude-estimation variants of the Hadamard test, which reduce query complexity to $\mathcal{O}(1/\epsilon)$ at the cost of deeper controlled circuits \cite{Brassard2002QAE}.

Beyond measurement overhead, our results indicate that extended-time accuracy is primarily limited by reference-state choice: selecting states solely to maximize overlap with either the initial state or a short-time evolved state is generally insufficient, motivating more dynamics-aware selection criteria capable of achieving reliable long-time evolution with fewer Trotterization steps and reduced quantum resources. In particular, adaptive reference-state selection strategies, such as approaches based on Krylov residuals, or on optimizing the sampling time according to the spectral information content of the measured bitstring distribution, represent important open directions for improving the robustness and scalability of the framework.

\section{Summary and Conclusions \label{sec:summary_and_conclusions}}
We introduced a quantum-computer-aided framework for simulating and characterizing interacting spin-defect systems in solids, with the explicit aim of supporting the design, benchmarking, and optimization of quantum-technology platforms under experimentally relevant conditions. The approach is built around a general ESR many-body Hamiltonian incorporating zero-field splitting, Zeeman and hyperfine interactions, dipolar spin--spin coupling, and an effective spin--boson model of phonon-induced decoherence. To make such models tractable on NISQ and early fault-tolerant hardware, we combined (i) Gray-encoded qudit-to-qubit mappings for efficient operator representations, (ii) qubit-wise commuting (QWC) aggregation to reduce circuit depth and gate counts in Trotterized primitives, and (iii) the multi-reference selected Quantum Krylov Fast-Forwarding (sQKFF) algorithm to extend accessible dynamical timescales beyond what straightforward product-formula simulation would permit at comparable resources.

Using the NV$^{-}$ center in diamond as a benchmark testbed, we evaluated three representative configurations. Across these cases, the framework enabled the computation of operationally meaningful dynamical quantities, including autocorrelation functions, microwave absorption spectra relevant to ODMR, and the $\ell_{1}$-norm of coherence as a practical indicator of coherence decay. For Configurations 1 and 3, the resulting autocorrelation functions agreed closely with QuTiP reference simulations for tens of nanoseconds, while the absorption spectrum and coherence trends for Configuration 1 were reproduced faithfully across a range of Trotterization thresholds and reference-state counts. Resource estimates for the Hadamard-test circuits required to construct the Krylov overlap and Hamiltonian matrices, under the Toeplitz structure assumption, showed that QWC aggregation yields systematic savings of approximately $18$--$30\%$ in gate counts and circuit depth, with reductions remaining stable as the Trotterization error is tightened.

A central conclusion of our numerical study is that, at fixed hardware cost, long-time accuracy in sQKFF is governed primarily by the choice and number of reference states, rather than by Trotterization error alone. This effect is especially evident in the larger and more strongly interacting Configuration 2, where measurement probability mass is distributed over many bitstrings, making naive reference selection less effective. These observations motivate the development of more physically informed and dynamics-aware strategies for reference selection and sampling time, which could substantially improve accuracy without proportionally increasing quantum resources.

Overall, this work establishes a practical blueprint for leveraging quantum computers as tools for quantum device engineering in the solid state: not only to reproduce short-time dynamics, but to access longer operational timescales and extract experimentally relevant observables from realistic ESR spin-defect models. While realizing the full workflow on hardware will require careful accounting of measurement overhead, noise sensitivity, and subspace conditioning, the results here indicate a clear path toward scalable simulations that can guide the comparison and optimization of candidate spin-defect technologies. In this sense, the proposed methodology provides a concrete step toward using quantum computation to accelerate the discovery, validation, and refinement of next-generation quantum materials and spin-based quantum devices.

\section*{Key software packages}
The framework was implemented using PennyLane v0.42 and Catalyst v0.12 \cite{PennyLane}, with numerical results compared against approximated solutions obtained with QuTiP v5.2.1 \cite{qutip}. QuTiP solutions were computed using the \verb|sesolve| function with \verb|adams| method for ODE integration, absolute tolerance $10^{-5}$, and relative tolerance $10^{-6}$.

\begin{acknowledgments}
The authors acknowledge the Quantum Open Source Foundation Mentorship Program for enabling the participation of students interested in quantum technologies in research projects such as the one presented here. \par
We extend our thanks to Kevin Ferreira, Yipeng Ji and Paria Nejat of the LG Electronics Toronto AI Lab and to Sean Kim of LG Electronics, AI Lab, for their ongoing support of our research.
\end{acknowledgments}

\bibliography{ms.bib}

\end{document}